# Light-Addressable Smart Nanostructures via Resonant Nanoheating.

*Victor Tabouillot[1]*, Douglas Murad[1], Rahul Kumar[1], Paula L. Lalaguna[1], Maryam Hajji[1], Affar Karimullah[1], Nikolaj Gadegaard[2], Aurélie Malfait[3], Patrice Woisel[3], Graeme Cooke[1], and Malcolm Kadodwala[1]**

[1] School of Chemistry, Joseph Black Building, University of Glasgow, Glasgow, G12 8QQ, UK

[2] School of Engineering, Rankine Building, University of Glasgow, Glasgow G12 8LT, U.K

[3]Univ. Lille, CNRS, INRAE, Centrale Lille, UMR 8207 - UMET - Unité Matériaux et Transformations, F-59000 Lille, France

## Abstract

Selective spatial control of chemical reactions at the level of individual nanostructures remains a significant challenge. We introduce a light-activated system that combines plasmonic gold nanorods with a poly(N-isopropylacrylamide) monolayer to gate surface reactivity based on each rod's geometry under optical illumination. Laser excitation tuned to a rod's plasmon resonance and polarization collapses the polymer into a compact shell on that rod, blocking reactive head groups and creating a long lived kinetically trapped inert state stable for days.

During this interval, orthogonal chemical transformations can be performed on adjacent, unilluminated rods without interference. Subsequent diffusion-limited rehydration restores the swollen brush conformation and renews surface activity, effectively erasing the chemical memory. Numerical simulations based on real nanorod geometries confirm that switching selectivity follows the rods' absorption profiles. This mask-free, fully reversible strategy turns passive polymer films into dynamic chemical interfaces, offering a route to high-resolution patterning and on-demand control of nanoscale reactions for electronic and sensing applications.

## Introduction

Selective functionalization of plasmonic nanostructures underpins advances in biosensing[1, 2], optoelectronics[3], photocatalysis[4] and photonic device engineering[5]. To integrate molecular function into complex architectures, one must not only attach chemical moieties at the nanoscale but also address individual structures differently across a substrate. Conventional "top-down" patterning techniques—inkjet printing[6], dip-pen nanolithography[5], and direct laser writing[7]—offer sub-100 nm resolution in ideal cases, but rely on serial processing, masks or tips, and often struggle with throughput, material compatibility or complex three-dimensional geometries[8, 9]

Thermoplasmonics provides a contactless, far-field route to localized heating via resonant light absorption by metallic nanostructures[10, 11]. This approach has driven a range of photothermal transformations and dynamic surface processes[12], yet remains agnostic to individual particle shape or aspect ratio. In other words, existing thermoplasmonic schemes cannot selectively address one nanoparticle or nanorod in an ensemble without beam shaping or near-field probes.

Here, we bridge that gap by coupling geometry-dependent plasmonic heating with the reversible collapse of a thermoresponsive poly(N-isopropylacrylamide) (p-NIPAM) self-assembled monolayer. By tuning the wavelength and polarization of a uniform far-field laser beam to each nanorod's longitudinal resonance, we collapse the polymer shell only on rods whose aspect ratio matches the excitation conditions, sterically blocking functional head groups for a programmable, multi-day "off" state. During this latency window, orthogonal chemistries can be performed on unilluminated rods in the same array; subsequent diffusion-limited rehydration restores polymer conformation and reactivity, effectively erasing chemical memory.

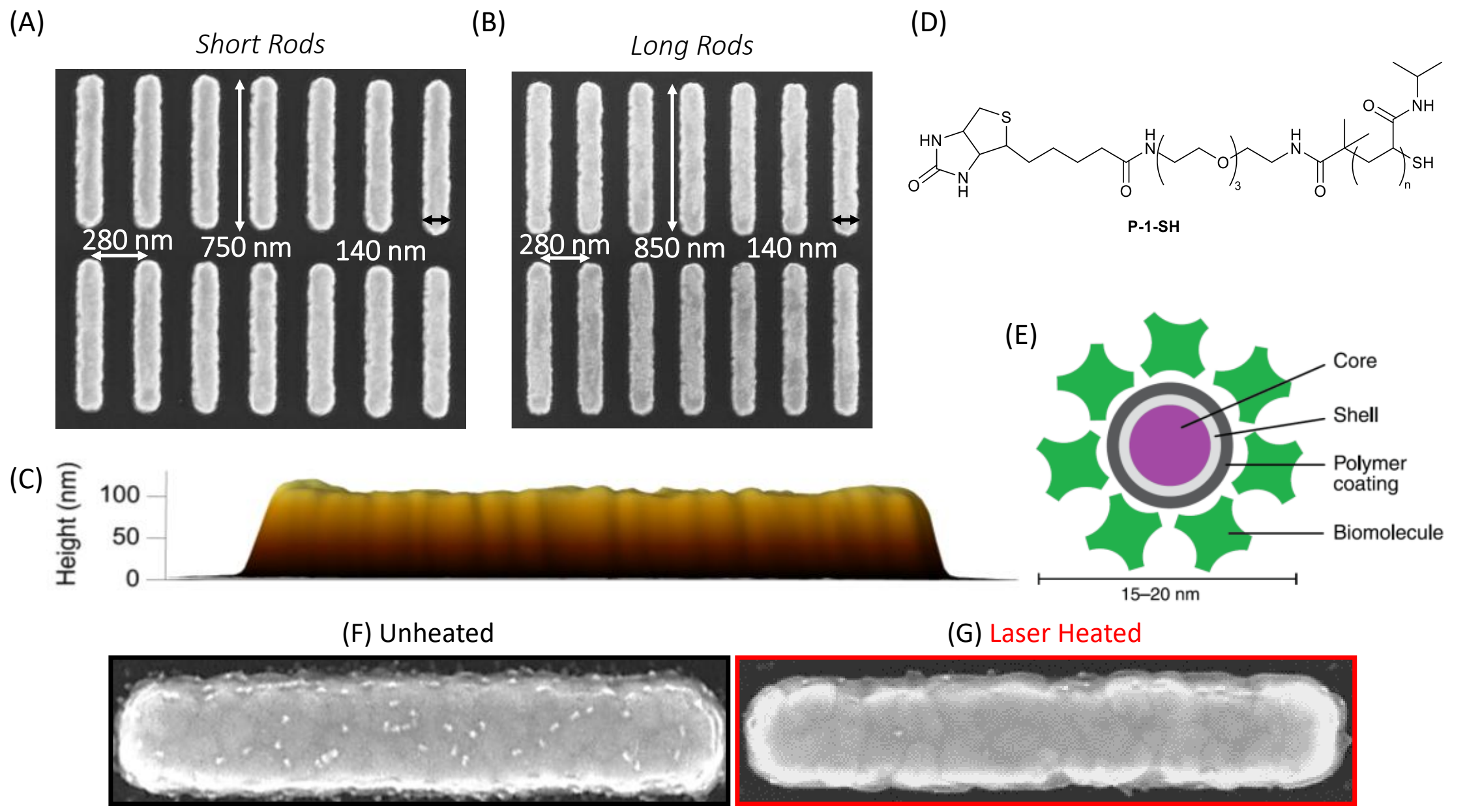

**Figure 1**. SEM of (a) 750 nm short nanorods array and (b) 850 nm long rods array. (c) AFM image of a rod from the side. (d) Thermal responsive polymer *p-NIPAM* (P-1-SH) molecule and (e) schematic of a QD streptavidin conjugate. SEM of (f) unilluminated rod after QDs functionalization and (g) after a 660 nm pulsed laser heating longitudinal polarization at a fluence of 255 mJ cm$^{-2}$.

Our mask-free strategy thus encodes reactivity in nanostructure geometry alone, requires no lithography or structured beams, and delivers single-structure selectivity with only far-field optics. Realistic electromagnetic and thermal simulations, based on experimentally derived geometries, confirm that switching correlates precisely with each rod's absorption profile. This new paradigm transforms static polymer coatings into dynamic, programmable chemical interfaces, opening routes to high-throughput, high-resolution patterning and smart-material memory at the nanoscale.

## Results and Discussion

**Nanostructures:** To explore aspect-ratio–selective functionalization, we fabricated 1 × 1 $mm^2$ arrays of plasmonic gold nanorods with two distinct aspect ratios by electron-beam lithography. Each array contained either "short" rods (~ 750 nm long) or "long" rods (~ 850 nm long), both with a transverse width of ≈ 140 nm and a height of ≈ 110 nm, arranged on a silicon substrate at a 280 nm pitch. The lateral dimensions were confirmed by scanning electron microscopy (SEM), and the heights by atomic force microscopy (AFM) (Figure 1A–C).

We characterized the optical response of the unfunctionalized nanorods in buffer using reflectance spectroscopy. Spectra were recorded with the polarization aligned along the long axis (0°) and the short axis (90°) of the rods; data for the short and long arrays (Supplementary Information). In the 0° spectra, enhanced reflectance peaks appear at ~ 680 nm for short rods and ~720 nm for long rods. By contrast, the 90° spectra of both arrays are nearly identical—which is expected, since the rods share the same width—and exhibit a resonance of increased transmission at ~700 nm. The line shape of the reflectance spectra can be replicated using electromagnetic (EM) numerical simulation (Comsol Multiphysics) (Supplementary Information).

**Surface functionalisation**: A self-assembled monolayer of thiol-terminated, biotin-functionalized poly(N-isopropylacrylamide), Figure 1 D, was formed on gold nanorods by immersion in polymer solution for 24 h, a treatment known to produce saturated, well-ordered monolayers within a day on Au surfaces[13]. Unlike surface-initiated polymer brushes (grafting-from)[14], which grow high-density chains in an extended conformation, this solution-phase adsorption (grafting-to) deposits preformed chains that maintain their solution-phase coil/helix structure at the interface. Consequently, the biotin head group is accessible to bind to streptavidin functionalised quantum dots (CD), Figure 1E. Successful SAM formation was confirmed by a 2–3 nm red shift in the plasmon resonance and an ≈10 nm increase in its linewidth (Supplementary Information). The red shift arises from the higher local refractive index at the metal interface introduced by the SAM[15], while the newly formed metal–chemisorbed adsorbate boundary enables chemical interface damping—additional nonradiative decay pathways that broaden the plasmon linewidth[16].

**Thermal properties of SAM:** The thermally responsive properties of p-NIPAM in solution are well established. It exhibits a lower critical solution temperature (LCST) of ~316 K[17], above which the polymer undergoes a reversible conformational collapse from an extended helical state to a compact globular form. The polymer used to form the SAM conformed to this behaviour as illustrated with optical transmission and dynamic light scattering measurements, (Supplementary Infomration). Below its LCST, p-NIPAM chains remain fully

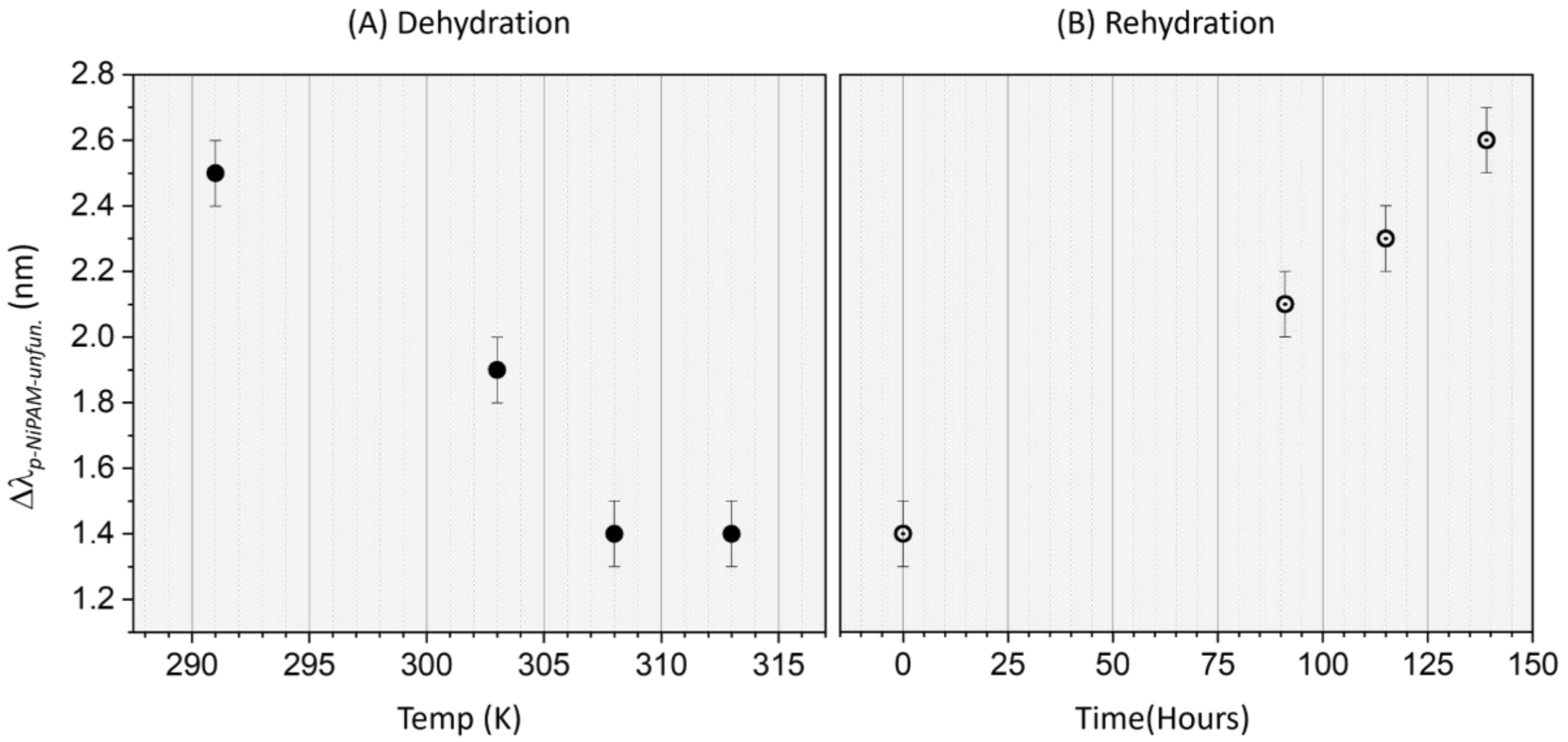

**Figure 2**. (A) Dehydration and collapse of the p-NIPAM SAM is monitored by measuring Δλp-NiPAM-unfun.as a function of the temperature of the surrounding buffer solution. (B) Rehydration monitored by measuring Δλp-NiPAM-unfun. as a function of time.

hydrated in a random-coil (helical) conformation and the solution is optically clear, transmitting nearly 100 % of incident visible light. As the temperature exceeds the LCST, the polymer collapses into dense, hydrophobic globules that form large aggregates and scatter light strongly, producing a sharp, drop in transmittance and increase in the hydrodynamic radius at the transition temperature

Thermal behaviour of the p-NIPAM SAM was evaluated by immersing substrates in buffer at each target temperature for 15 minutes and recording reflectance spectra. After cooling to room temperature, spectra were acquired within 30 minutes. The resulting plasmonic resonance shift ($\Delta\lambda$), relative to an unmodified array, was plotted versus temperature, Figure 2A. Heating to 308 K induced a 1.1±0.2 nm decrease in $\Delta\lambda$, which persisted up to 313 K. To test reversibility, samples were returned to buffer at room temperature for up to 138 hours, with spectra collected periodically, Figure 2B. Over this period, $\Delta\lambda$ gradually recovered, reaching its original value after 138 hours.

Because the temperature at which $\Delta\lambda$ decreases closely matches the expected LCST of p-NIPAM, we attribute this shift to polymer collapse into a globular conformation. However, our results differ notably from earlier reflectance-based studies of p-NIPAM–functionalized gold plasmonic nanoparticles: in those studies, collapse led to a red shift—indicative of an increased effective refractive index in the near field[18-20], whereas here we observe a blue shift, signalling a reduced effective index. We believe this can be reconciled by noting that the nanorods studied here are significantly larger than the previously examined nanoparticles (< 100 nm). This larger size strongly modifies the EM near field: small nanoparticles (< 100 nm) support fields that decay within ~10 nm of the surface, but larger structures resonating at longer wavelengths generate near fields that extend significantly farther (validated by numerical simulations; Supplementary Information). Thus, for nanorods, the near field extends beyond the collapsed polymer layer, so collapse replaces higher-index polymer with lower-index aqueous buffer, lowering the average refractive index in the near-field volume and producing a small blue shift. In contrast, for small nanoparticles the near field remains within the collapsed globule, which contains less water and thus increases the local refractive index, yielding a red shift..

The kinetics of collapse/rehydration of p-NIPAM SAMs have not been reported. In contrast, p-NIPAM–coated gold nanoparticles recover from collapse in ∼100 ns following laser flash photolysis experiments[19]. Densely grafted p-NIPAM brushes on planar substrates reswell fully within seconds to minutes: QCM-D studies show complete rehydration in 1–10 s[21], and interferometric measurements confirm recovery in ∼100–200s [22]. We attribute the multi-day equilibration of SAMs to several factors: (i) low grafting density (<1 chain·nm$^{-2}$) inherent to thiol–gold monolayers, which causes chains to collapse into compact, substrate-pinned globules with minimal free volume[14]; (ii) strong thiolate–gold anchoring, which prevents chain desorption and limits mobility[23]; and (iii) enhanced lateral van der Waals interactions in low-density SAMs, which raise the energy barrier to chain re-extension by orders of magnitude compared to brushes[24]. Together, these features convert an inherently fast LCST-driven transition into a diffusion-limited, kinetically trapped state that only returns to equilibrium over days.

**Laser driven process:** Thermoplasmonic activation of the polymer layer was achieved using a nanosecond pulsed laser (5 ns pulse width, 20 mW average power, 255 mJ cm$^{-2}$ fluence), directed at normal incidence through a 10× objective to a ∼1 mm diameter spot. By tuning the laser wavelength and polarization, we selectively excited the plasmonic resonances of nanorod arrays with specific dimensions, producing localized heating through nonradiative decay [10]. Arrays were exposed for 2 minutes to various combinations of wavelength (660–1000 nm) and polarization (longitudinal or transverse relative to the nanorod long axis).

To probe the functional state of the polymer layer after laser exposure, the samples were incubated with streptavidin-functionalized quantum dots. These bind only to exposed biotin groups, providing a direct visual readout of surface chemical activity[25]. SEM imaging revealed substantial differences in quantum dot binding between irradiated and unexposed arrays. As shown in Figure 1F–G, unheated nanorods exhibited dense QD coverage, whereas irradiated rods displayed significantly reduced binding, consistent with thermally induced polymer collapse and biotin deactivation. The contrast in QD binding underpins the form-factor-selective functionalization strategy explored in the following sections.

The thermally induced switching of the p-NIPAM layer was first monitored via reflectance spectroscopy (Supplementary Information), exploiting the sensitivity of the nanorod plasmonic resonance to changes in the surrounding dielectric environment [26] [1]. For uncoated arrays in water, the longitudinal (0°) reflectance spectra revealed resonance peaks centered at ~680 nm for short nanorods and ~720 nm for long nanorods. Following adsorption of the p-NIPAM monolayer, a red shift of 2–3 nm was observed for both structures[27].

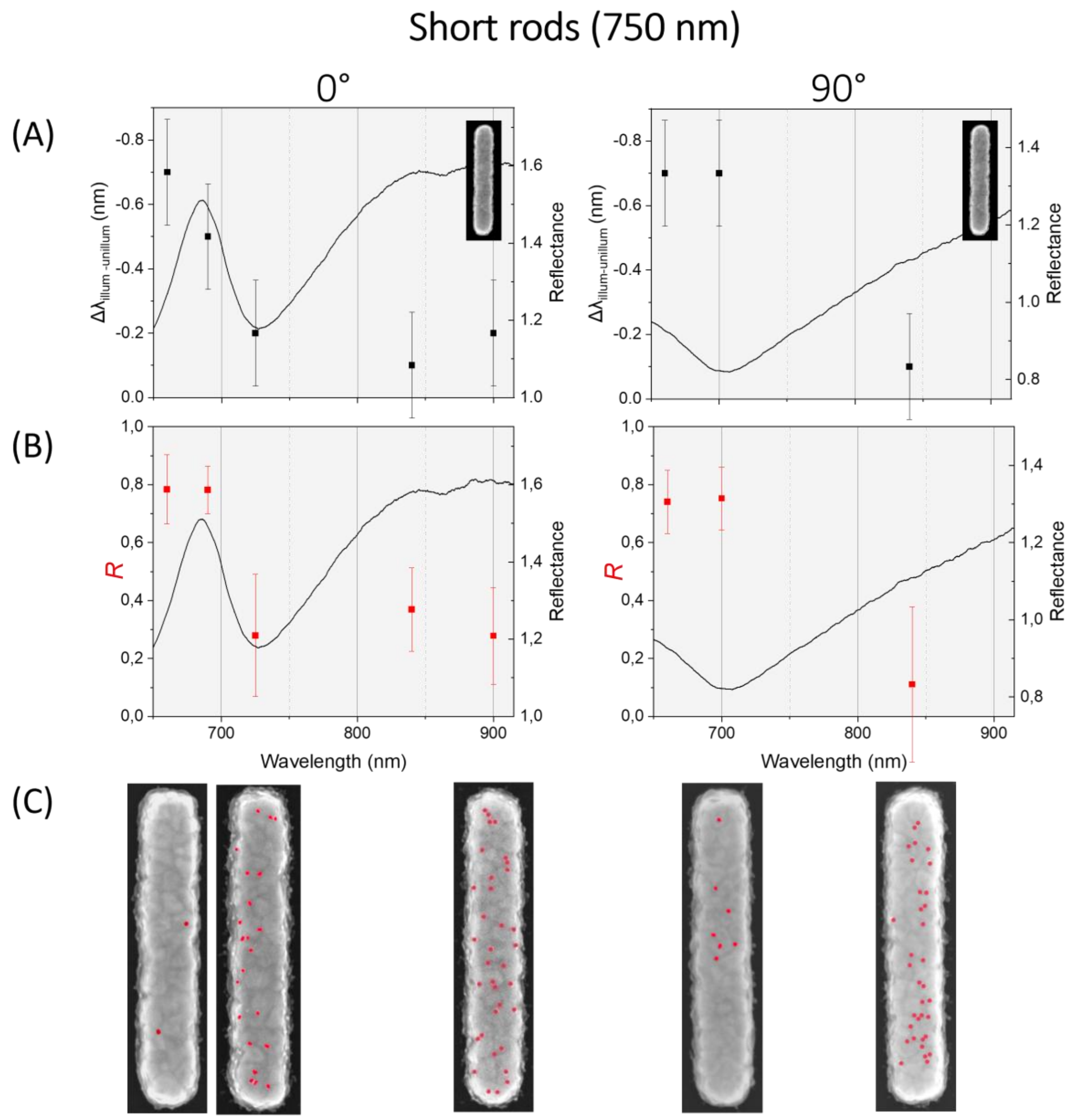

**Figure 3**. Short rods data of (A) reflectance shift $\Delta\lambda_{illum-unillum}$ as a function of the wavelength of the pulsed laser, for both 0° and 90° laser polarizations on the left and right column respectively (shown by the red arrow relative to a rod on the top right of each plot). The reflectance spectra corresponding to the laser polarization were overlayed on each panel. The error bar represents the standard deviation of 5 spectra. (B) shows the R factor in red, relative to all the laser wavelengths used at both polarization angles. The error bar is the standard deviation from the QDs count of 30 different rods. (C) SEM images of a rod corresponding to the pulsed laser illumination indicated by the purple or yellow arrows, with colorized QDs in red (the white scall bar on the far-left image represents 100 nm).

After laser irradiation, a wavelength-dependent blue shift was observed in the plasmonic resonance peak position, indicating a local reduction in refractive index consistent with the thermally induced collapse of the p-NIPAM layer [18]. The magnitude of the shift varied with both laser wavelength and polarization, with more pronounced shifts occurring when the laser excitation was close to the resonance of the nanorod array. This behaviour is summarised in Figures 3A and 4A, which plot the change in resonance wavelength ($\Delta\lambda$) as a function of laser wavelength for short and long nanorods, respectively.

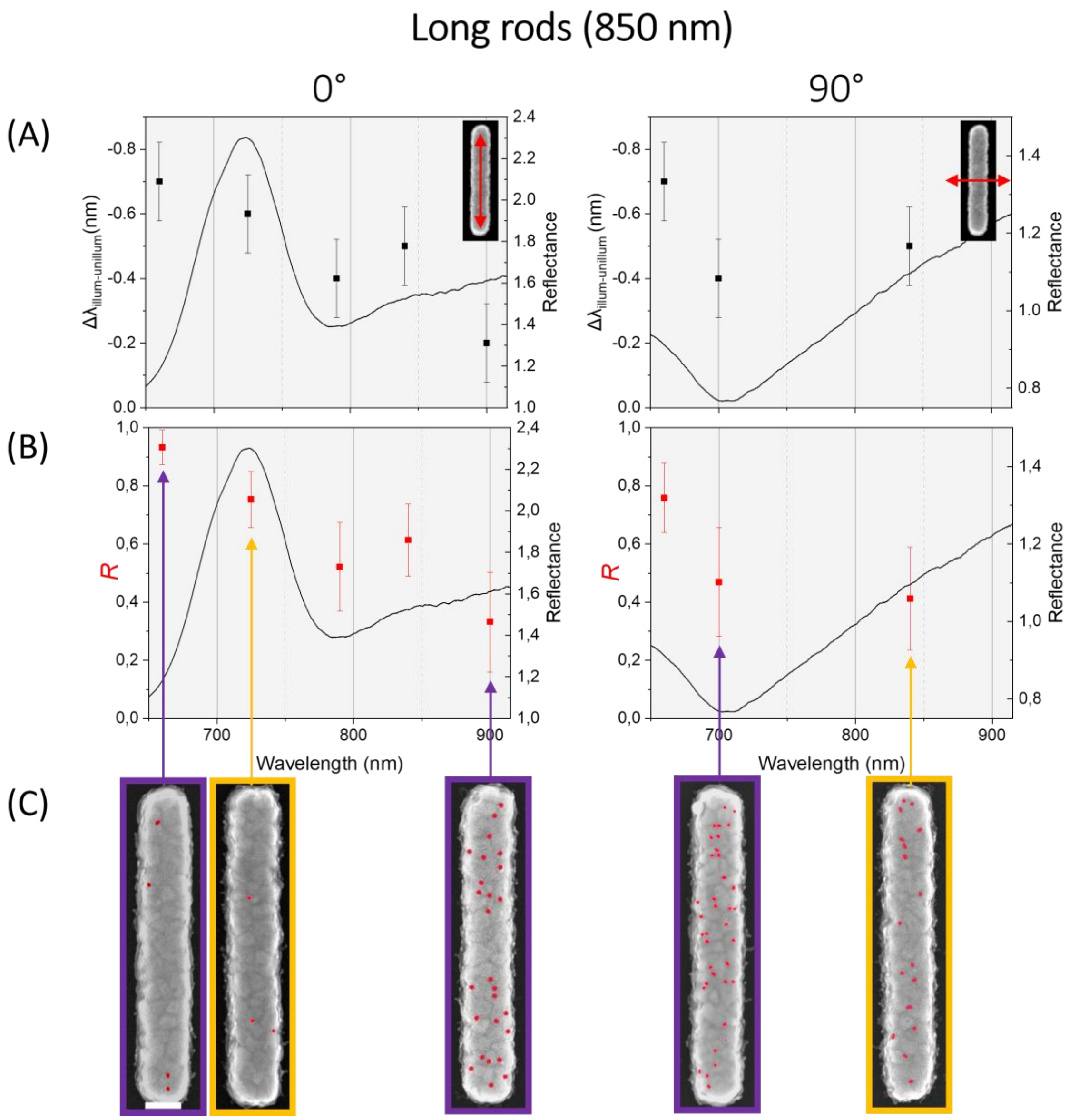

**Figure 4** Long rods data of (A) reflectance shift $\Delta\lambda_{illum-unillum}$ as a function of the wavelength of the pulsed laser, for both 0° and 90° laser polarizations on the left and right column respectively (shown by the red arrow relative to a rod on the top right of each plot). The reflectance spectra corresponding to the laser polarization were overlayed on each panel. The error bar represents the standard deviation of 5 spectra. (B) shows the R factor in red, relative to all the laser wavelengths used at both polarization angles. The error bar is the standard deviation from the QDs count of 30 different rods. (C) SEM images of a rod corresponding to the pulsed laser illumination indicated by the purple or yellow arrows, with colorized QDs in red (the white scall bar on the far-left image represents 100 nm).

To directly assess the chemical functionality of the polymer-coated surfaces, we quantified the number of streptavidin-conjugated quantum dots bound to each nanorod using SEM imaging (Supplementary Information). A reduction factor, R, was defined as the ratio of the average number of QDs bound to irradiated rods compared to unirradiated controls. An R value of 1 indicates complete suppression of binding (i.e., full polymer collapse), while $R \approx 0$ signifies no change from unheated arrays. The variation of R with laser parameters is shown in Figures 3B and 4B.

This analysis revealed a strong dependence of QD binding on both nanorod geometry and laser excitation conditions. For example, under 725 nm longitudinal (0°) illumination, long rods showed a marked reduction in QD coverage ($R \approx 0.9$), whereas short rods remained largely functional ($R \approx 0.1$). The inverse behaviour was observed at 700 nm with transverse (90°) polarization, where QD binding was suppressed on short rods but retained on long rods. Figures 3C and 4C present representative SEM images of the corresponding QD binding patterns under these irradiation conditions, providing direct visual confirmation of the quantitative trends shown in Figures 3B and 4B.

These results confirm that by tuning the excitation wavelength and polarization, it is possible to selectively activate or suppress surface chemistry based on nanostructure form factor. Crucially, the trends in plasmonic blue shift ($\Delta\lambda$) and QD reduction factor (R) exhibit strong correlation: both reflect the same underlying resonance-dependent heating mechanism. This is clearly evidenced by the agreement between Figures 3A and 3B for short rods and Figures 4A and 4B for long rods, where maxima in $\Delta\lambda$ align closely with the conditions that yield maximum suppression of QD binding. These correspondences validate the use of far-field optical measurements as a quantitative and predictive proxy for the functional state of surface chemistry.

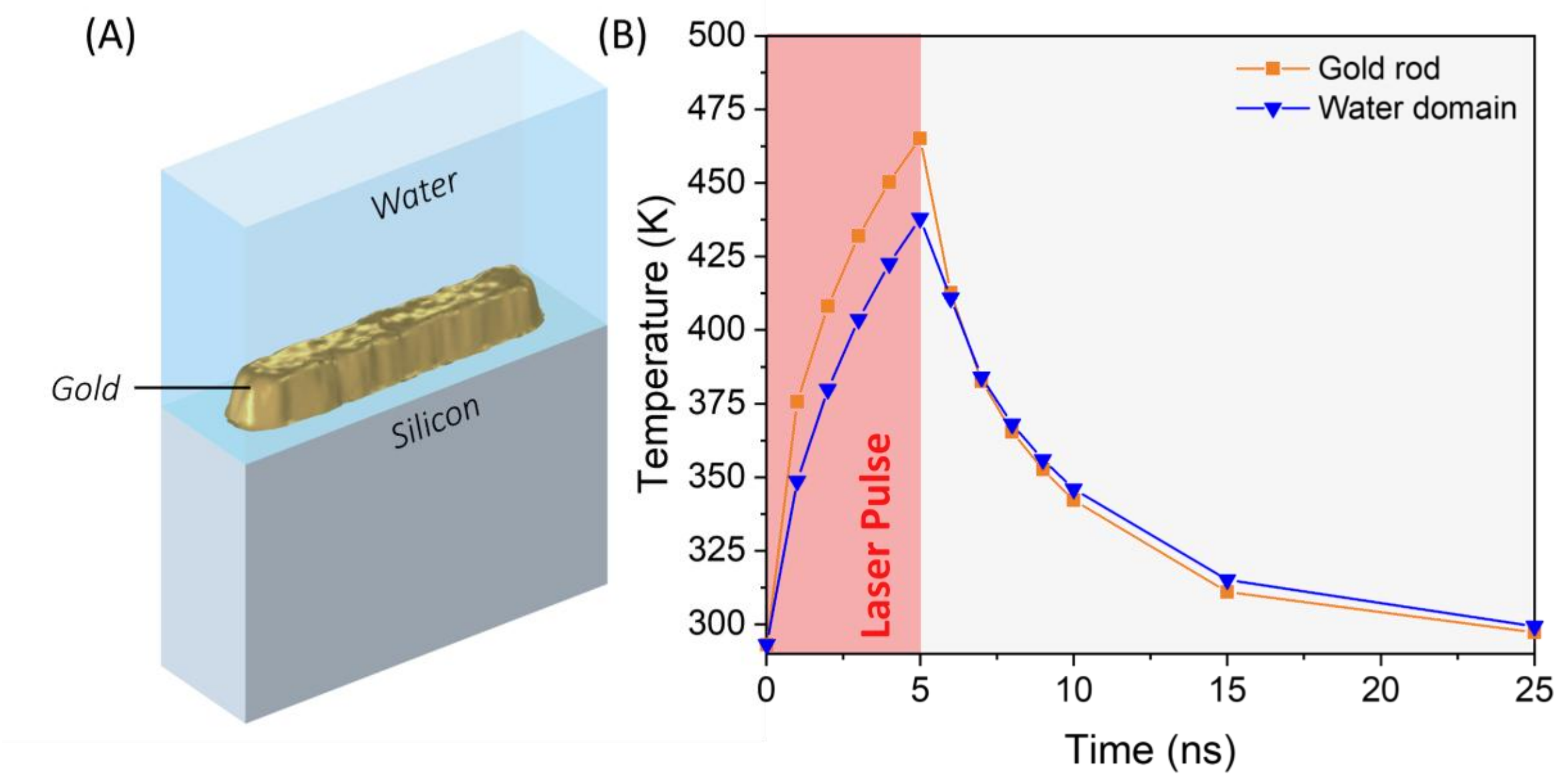

**Figure 5**. (A) COMSOL model of a short gold rod obtained by AFM imaging and (B) the average temperature as a function of time in the short rod (orange) and inside a 10 nm water domain near the surface (blue). The 5ns laser was set at a 0° polarization angle from the rod, and at $\lambda = 690$ nm.

The thermoresponsive behaviour of the p-NIPAM layer was found to be gradual and temperature-dependent, consistent with previous studies on polymer collapse dynamics [19, 28]. As a result, partial switching and intermediate QD densities were observed under some irradiation conditions, enabling graded control over surface reactivity. These findings underscore the tunability of the system and demonstrate that far-field optical inputs can programmably modulate chemical functionality at the level of individual nanostructure arrays across macroscopic areas.

To understand the mechanism driving the observed selective switching behaviour, we conducted finite-element simulations of electromagnetic absorption and transient heat transfer using COMSOL Multiphysics. The model geometry was constructed from AFM topographic data of the fabricated nanorods, rather than idealised analytical shapes. This realistic geometry captures fabrication-related features such as sidewall tapering, curvature, and edge asymmetries, enabling more accurate prediction of both plasmonic absorption and heat localisation. The COMSOL simulation geometry is shown in Figure 5A, and represents a methodological refinement over conventional simulations based on idealised structures.

The thermal response of the system was calculated using the transient heat diffusion equation:

$$\rho C_p \frac{\partial T}{\partial t} - \nabla.(k \nabla T) = Q \qquad (1)$$

where $\rho$ is the density, $C_P$ the specific heat, $k$ the thermal conductivity, $T$ the temperature, and $Q$ the electromagnetic heat source term (power density). This equation (eq 1) was solved using time-dependent conditions matching the experimental laser pulse duration (5 ns).

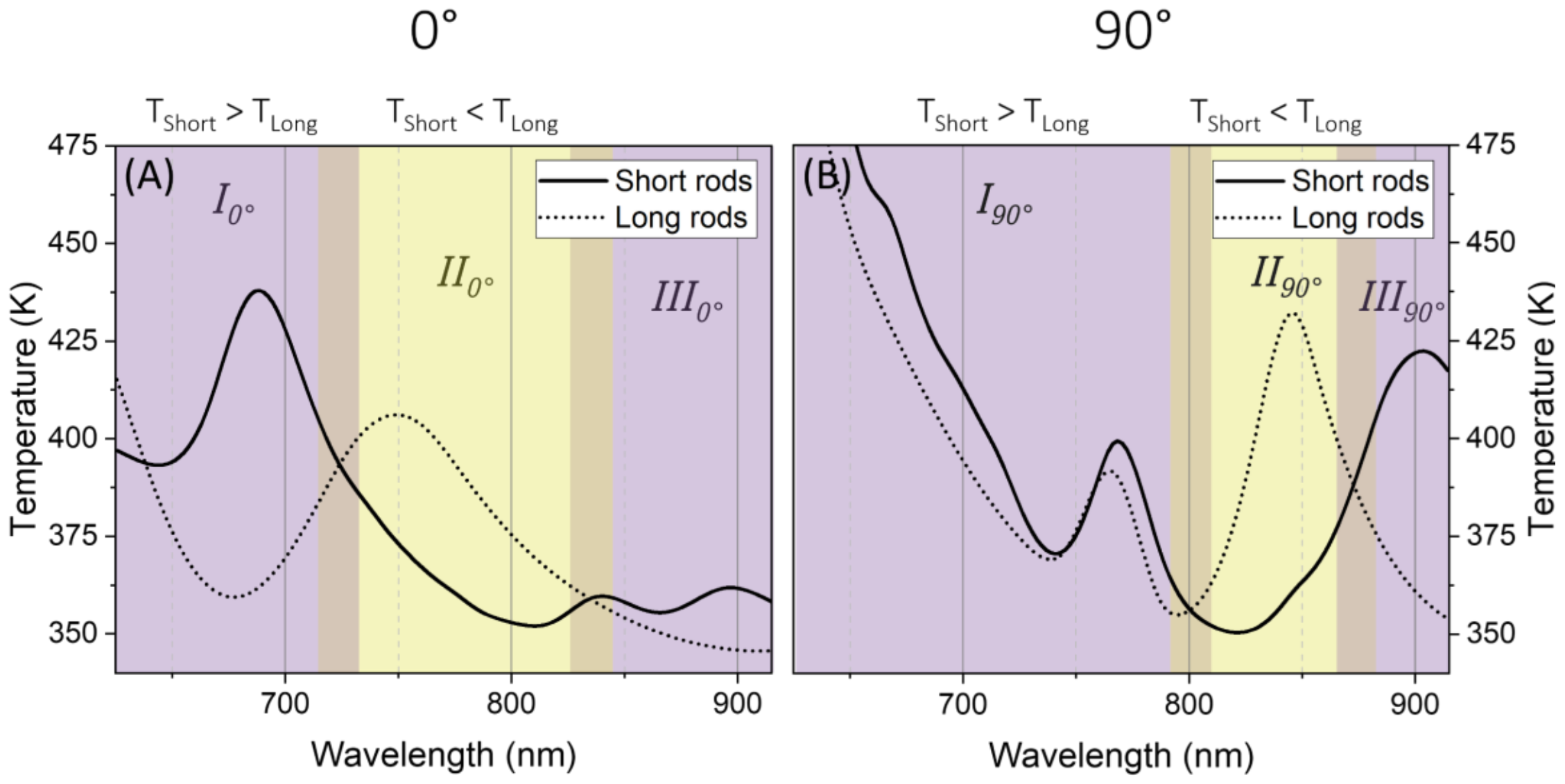

**Figure 6** Average temperature after a 5 ns laser pulse in a 10 nm water area around a short rod (solid line) and a long rod (dotted line) as a function of the laser wavelength. The left panel displays the results for longitudinal polarization, while the right panel is for the corresponding transversal polarization. The purple zone indicates the wavelength range where the temperature for the short rod case is higher than for the long rod, and the yellow zone represents the opposite scenario.

The power density $Q$ was derived from the simulated electric field distributions using:

$$Q = \frac{\omega}{2} Im\{\boldsymbol{P}^{*}.\boldsymbol{E}\} \qquad (2)$$

where $\boldsymbol{E}$ is the electric field and $\boldsymbol{P}$ the polarisation field. In isotropic media, this simplifies to:

$$q = \frac{\omega}{2} Im(\varepsilon(\omega))\varepsilon_0 |E|^2 \qquad (3)$$

Equations 2 and 3 were used to quantify the absorption profile within the nanorods, with optical constants taken from Johnson and Christy [29].

The temporal evolution of temperature during a single 5 ns laser pulse is shown in Figure 5B. The gold nanostructure reaches its peak temperature rapidly during the pulse, while the surrounding water heats more slowly and to a lower maximum. This transient mismatch indicates that the p-NIPAM layer at the rod–solution interface experiences a brief thermal spike sufficient to trigger collapse, while the bulk remains below the LCST (~316 K)[17].

The simulated peak temperatures for short and long rods under varying excitation conditions are summarised in Figure 6, which identifies three thermal regimes (Regions I–III, subscripts label 0° / 90°) based on the relative heating of the two geometries. Notably, the simulations show that peak temperatures exceed the LCST of p-NIPAM across all wavelengths and polarisation conditions. However, switching is only observed experimentally under resonance conditions — and under these conditions, complete switching is achieved, as evidenced by the full suppression of quantum dot binding (R ≈ 1). In both Region I and Region III, the short rods are hotter than the long rods, while in Region II, the long rods are hotter than the short rods. This inversion enables selective switching of the long rods. Region II spans different wavelength ranges for 0° and 90° polarisation, meaning that polarisation can be used to selectively target nanorod geometries for functionalisation at a fixed wavelength. This provides a mechanistic explanation for the polarisation-dependent switching observed experimentally (Figures 3 and 4).

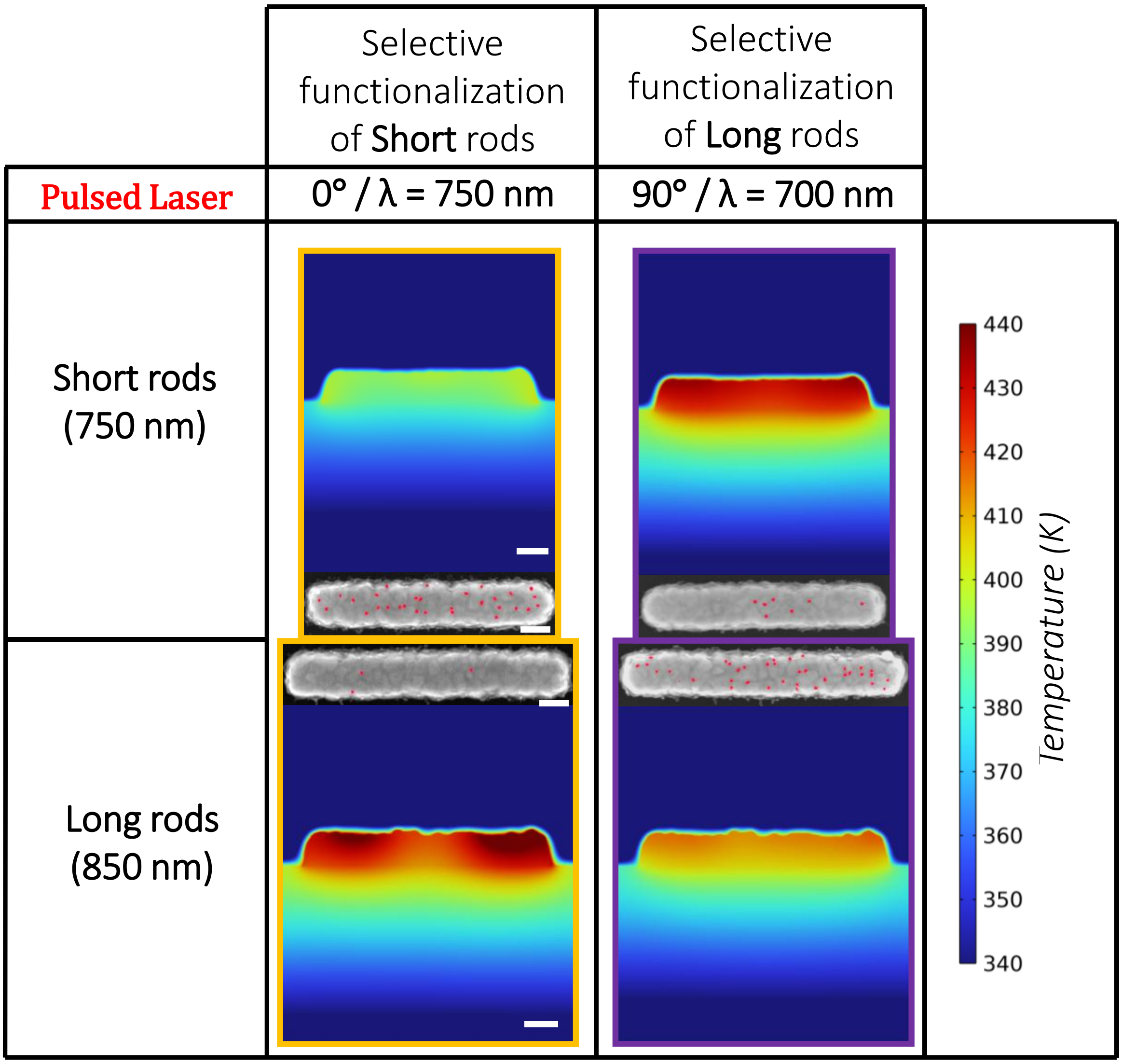

**Figure 7** Simulated heat maps obtained after 5ns illumination and experimental SEM images of both rods' lengths showing selective QDs functionalization at different laser wavelength and polarization angle. The SEM on the left panel (yellow) were obtained for a laser illumination of λ = 725 nm at 0°, while the SEM on the right panel (purple) were taken for a λ = 700 nm 90° pulsed laser illumination. The white scall bars on the left panel represent 100 nm.

This apparent paradox — that switching is selective despite all simulated conditions exceeding the LCST — can be resolved by considering the non-equilibrium nature of pulsed laser heating.

The LCST defines a thermodynamic transition temperature under slow, uniform heating[17], but in our system, thermal excitation occurs over nanosecond timescales and is followed by rapid cooling. Under these conditions, polymer collapse requires not only exceeding the LCST but also sufficient residence time above it for dehydration and conformational reorganisation. Off-resonance conditions may briefly exceed 316 K, but the combination of shorter duration and smaller effective volume reduces the likelihood of collapse. This kinetic explanation is consistent with previous reports of rate-dependent behaviour in p-NIPAM films [19]. To further probe the threshold conditions for switching, this experiment was replicated at half the laser power (10 mW average; see Supplementary Information). The results showed that QD binding was only marginally reduced for the combinations of 690 nm/0° on short rods ($R = 0.2 \pm 0.15$), and 660 nm/90° on both short and long rods ($R = 0.3 \pm 0.17$ and $R = 0.26 \pm 0.2$ respectively). This indicates that only these resonance-matched conditions raised the p-NIPAM layer above its LCST, even under reduced power, further confirming the selectivity of resonant nanoheating.

The spatial temperature distribution at the end of the pulse is shown in Figure 7, which confirms that heating is strongly confined to the nanostructure and its immediate surroundings. Most of the heat dissipates vertically into the silicon substrate, with minimal lateral diffusion into the surrounding fluid. This spatial confinement explains the lack of thermal cross-talk between adjacent structures and supports the interpretation of form-factor-specific switching under far-field illumination.

Together, these simulations provide a quantitative, time-resolved framework for interpreting the experimental results. The use of realistic nanostructure geometry, transient thermal modelling, and polarisation-resolved absorption analysis confirms that resonance-enhanced, structure-specific nanoheating enables programmable switching of surface chemistry using only optical input parameters.

## Conclusion

This study establishes a new paradigm in light-addressable, geometry-specific chemical gating at the nanoscale. By exploiting the aspect-ratio-dependent plasmonic absorption of gold nanorods, we achieve single-structure selectivity using only far-field optical inputs—no lithography, masks, or structured beams required. Resonant laser excitation collapses a thermoresponsive p-NIPAM SAM on targeted rods, generating a metastable “off” state that endures for longer than a day. This extended latency enables orthogonal chemical transformations on unilluminated structures before spontaneous rehydration restores reactivity. Realistic numerical simulations validate that switching correlates precisely with each rod’s resonant profile. Our platform thus converts passive polymer coatings into programmable chemical interfaces that encode, store, and reset surface reactivity through nanostructure geometry, offering a sexy, high-impact route to dynamic patterning and memory in smart materials..

## Experimental (Materials and Methods)

**Nanostructure Fabrication:** Gold plasmonic nanorod arrays were fabricated on silicon substrates using electron-beam lithography (EBPG 5200) on a positive PMMA resist. Following development, a 5 nm Ti adhesion layer and 100 nm of gold were deposited via thermal evaporation (Plassys MEB-400s), and lift-off was performed using a bi-layer process. Two distinct nanorod lengths (750 nm and 850 nm) were patterned into separate square arrays (1 mm × 1 mm).

**Polymer Functionalisation**: Poly(N-isopropylacrylamide) (p-NIPAM) functionalised with biotin-terminated thiol side groups was synthesised as described previously.[15] After oxygen plasma treatment (80 W, 2 min), substrates were immersed in 0.5 g/mL p-NIPAM in PBS buffer (pH 7.4) for 24 h, then rinsed with PBS to remove unbound polymer.

**Laser Nanoheating**: Nanostructures were irradiated using a nanosecond pulsed laser (Opotek Radiant SE 532 LD, 5 ns pulse width, 20 Hz repetition rate). Beam polarisation was controlled with two Glan-Laser polarisers, and fluence was tuned to 127–255 mJ $cm^{-2}$. The beam was directed normally onto the sample through an Olympus 10× objective (0.30 NA) to yield a 1 mm diameter illumination spot. Arrays were exposed for 2 min. Beam alignment and focus were monitored using a Blackfly USB3 camera. Beam power at the sample plane was verified using a calibrated thermal sensor (Thorlabs S425C).

**Quantum Dot Labelling**: After laser treatment, reflectance spectra were acquired, and samples were incubated overnight in a dark environment with a PBS solution containing streptavidin-functionalised quantum dots (14 µg/mL; Qdot 705, Invitrogen). Unbound QDs were removed by PBS rinsing.

**Optical Characterisation**: Reflectance spectra were measured with a custom-built setup[30], using a tungsten halogen source, Glan–Thompson polarisers, a 50:50 beam splitter, and an Ocean Optics USB4000 spectrometer. Spectra were recorded at longitudinal (0°) and transverse (90°) polarisation angles by comparing nanorod reflectivity to a bare silicon background.

**Imaging and Morphology**: Atomic force microscopy (AFM; Bruker Dimension Icon) was used to determine nanorod height. Scanning electron microscopy (SEM; Hitachi SU8240) provided lateral dimensions and pitch. QDs were visualised using secondary electron contrast between gold and semiconductor components.

**Numerical Simulations**: Finite-element simulations were performed using COMSOL Multiphysics v6.0. The Wave Optics module was used to calculate local field intensities and reflectance spectra under normally incident, linearly polarised illumination. The gold permittivity was taken from Palik's optical constants.[16] Periodic boundary conditions replicated the array, and perfectly matched layers minimised boundary reflections.

The Heat Transfer module simulated thermal response to a 5 ns square laser pulse using time-dependent boundary conditions. Gold and dielectric material properties were sourced from COMSOL's built-in database. The model geometry was constructed from experimental AFM profiles to capture nanorod curvature and tapering. Temperature-dependent variation in optical and thermal properties was neglected. All simulations assumed steady-state fluid and substrate boundaries; heat transfer into water and silicon was modelled without convective flow.

## ASSOCIATED CONTENT

**Supporting Information**.

Details on sample fabrication; Reflectance spectra; SEM images; Simulated reflectivity

## AUTHOR INFORMATION

**Corresponding Author**

*malcolm.kadodwala@glasgow.ac.uk, 2604448t@student.gla.ac.uk, School of Chemistry, Joseph Black Building, University of Glasgow, Glasgow, G12 8QQ, UK.

**Author Contributions**

The manuscript was written through contributions of all authors. All authors have given approval to the final version of the manuscript.

**Funding Sources**

## ACKNOWLEDGMENT

Technical support from the James Watt Nanofabrication Centre (JWNC) and VT acknowledges the EPSRC for the award of a studentship.

## TABLE OF CONTENTS GRAPHIC

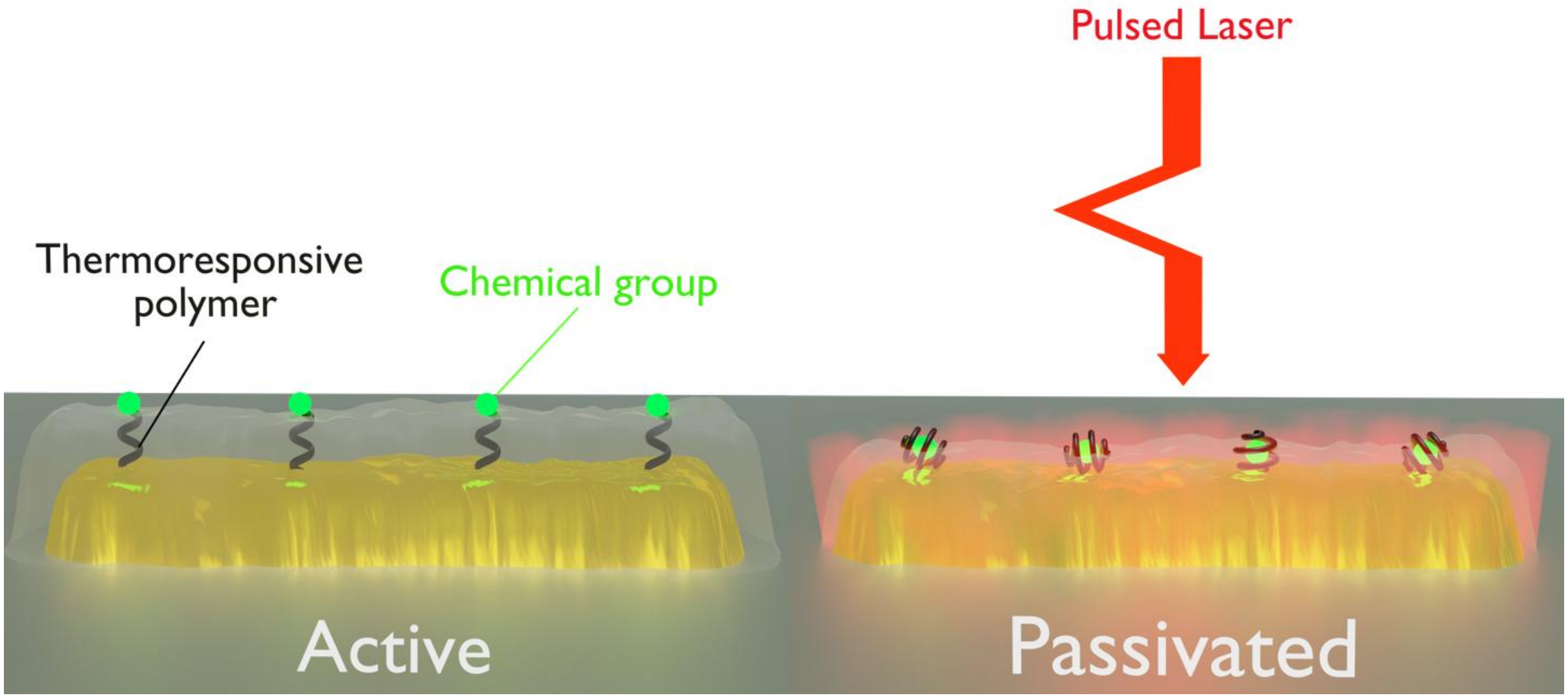